\documentclass[twocolumn,superscriptaddress,showkeys]{revtex4-1}
\usepackage{times}
\usepackage{bm}
\usepackage{graphicx}
\usepackage{amsmath}
\usepackage{amssymb}

\begin{document}
\title{Structural investigations of CeIrIn${_5}$ and
CeCoIn${_5}$ on macroscopic and atomic length scales}
\author{Steffen Wirth}
\affiliation{Max Planck Institute for Chemical Physics of Solids,
01187 Dresden, Germany}
\author{Yurii Prots}
\affiliation{Max Planck Institute for Chemical Physics of Solids,
01187 Dresden, Germany}
\author{Michael Wedel}
\affiliation{Max Planck Institute for Chemical Physics of Solids,
01187 Dresden, Germany}
\author{Stefan Ernst}
\affiliation{Max Planck Institute for Chemical Physics of Solids,
01187 Dresden, Germany}
\author{Stefan Kirchner}
\affiliation{Max Planck Institute for Physics of Complex Systems,
01187 Dresden, Germany} \affiliation{Max Planck Institute for
Chemical Physics of Solids, 01187 Dresden, Germany}
\author{Zachary Fisk}
\affiliation{University of California, Irvine, California 92697,
USA}
\author{Joe D. Thompson}
\affiliation{Los Alamos National Laboratory, Los Alamos, New
Mexico 87545, USA}
\author{Frank Steglich}
\affiliation{Max Planck Institute for Chemical Physics of Solids,
01187 Dresden, Germany}
\author{Yuri Grin}
\affiliation{Max Planck Institute for Chemical Physics of Solids,
01187 Dresden, Germany}

\begin{abstract}For any thorough investigation of complex physical
properties, as encountered in strongly correlated electron
systems, not only single crystals of highest quality but also a
detailed knowledge of the structural properties of the material
are pivotal prerequisites. Here, we combine physical and chemical
investigations on the prototypical heavy fermion superconductors
CeIrIn${_5}$ and CeCoIn${_5}$ on atomic and macroscopic length
scale to gain insight into their precise structural properties.
Our approach spans from enhanced resolution X-ray diffraction
experiments to atomic resolution by means of Scanning Tunneling
Microscopy (STM) and reveal a certain type of local features
(coexistence of minority and majority structural patterns) in the
tetragonal HoCoGa$_5$-type structure of both
compounds.
\end{abstract}

\keywords{heavy-fermion metals, structural organization, electron
density}
\date{\today}\maketitle

\section{Introduction}
Heavy-fermion metals typically contain magnetic 4$f$ or 5$f$
elements ({\it e.g.} Ce, Yb or U) which give rise to magnetism of
local moment character. At low temperatures, these magnetic
moments can effectively be screened by conduction electrons due to
the so-called Kondo effect \cite{Kondo}. The coupling of the free
electrons to the localized $f$ electrons can dramatically increase
the effective mass of the charge carriers which may reach up to
several hundred times the mass of a free electron. The properties
of these metals can often be described, according to Landau
\cite{Landau}, by considering quasi-particles made up of the
electrons and their interactions---instead of the mere
electrons---within the theory of a free electron gas. On the other
hand, in many of these materials an indirect exchange coupling
between the local magnetic moments is found (the so-called RKKY
interaction) which is also mediated---just like the aforementioned
Kondo interaction---via the conduction electrons. These two
interactions are in direct competition. The relative strength of
these two competing interactions can be tuned by experimental
parameters such as chemical substitution, pressure and magnetic
field. In case of this competition being adequately balanced a
quantum phase transition (QPT) at $T =$ 0 can be brought about by
a well-directed change of these experimental parameters
\cite{Doniach,SiSteg}.

Once the Kondo and RKKY interactions are well balanced additional,
smaller energy scales may play a decisive role. In fact, in many
cases superconductivity is observed in close proximity to a
quantum critical point (QCP), {\it i.e.} the point in phase space
at which a continuous QPT occurs \cite{Mathur}. Possibly,
superconductivity is one way of disposing the huge entropy
accumulated in the vicinity of a QCP. This concept has been
generalized \cite{Broun} such that possibly even in the
copper-oxide materials superconductivity might be related to a
hidden QCP. Here it should be noted that superconductivity for
which such a scenario is discussed is commonly considered to be of
unconventional nature, in a sense that the standard BCS theory
employing phonon-mediated Cooper pair formation \cite{BCS} cannot
be applied.

In this context the Ce$M$In$_5$ family of heavy-fermion compounds
offers an interesting playground \cite{SarThom,Tho12}. The
intricate interplay of superconductivity and magnetism is, {\it
e.g.}, manifested by the existence of superconductivity found in
CeCoIn$_5$ below the superconducting transition temperature
$T_{\rm c} \approx$ 2.3 K, and antiferromagnetic order in
CeRhIn$_5$ below the N\'{e}el temperature $T_{\rm N} \approx$ 3.7
K. Conversely, superconductivity with similar $T_{\rm c}$ is
observed in the latter compound by the application of pressure
\cite{Park} whereas neutron scattering experiments indicate strong
antiferromagnetic quasielastic excitations in the paramagnetic
regime of CeCoIn$_5$ \cite{Stock}. Moreover, the existence of a
field-induced QCP has been anticipated
\cite{pag03,bia03,sur07,zaum11,hu12,geg13}. Also, in
Cd-substituted CeCoIn$_5$ ({\it i.e.} CeCo(In$_{1-x}$Cd$_x$)$_5$
with $x \leq 0.01$) a microscopic coexistence and mutual influence
of the superconducting and antiferromagnetic order via identical
4$f$ states was inferred \cite{urb07,nair10}.

We report on structural investigations by both,
enhanced-resolution X-ray diffraction experiments as well as
atomically resolved Scanning Tunneling Microscopy (STM) on single
crystals of CeIrIn$_5$ and CeCoIn$_5$. These measurements indicate
the existence of a certain type of local structural features in
the tetragonal HoCoGa$_5$-type matrix that can directly be
visualized by STM. These features can be considered as patches of
the closely related TlAsPd$_5$ structure. They may also be related
to the apparent discrepancy in bulk $T_{\rm c} \approx$ 0.4 K and
resistive $T_{\rm c} \approx$ 1.2 K in CeIrIn$_5$
\cite{pet01,bia01,nair08}. Our results obtained by Scanning
Tunneling Spectroscopy (STS) confirm CeCoIn$_5$ to be a $d$-wave
superconductor \cite{davis13,yaz13} and indicate a precursor state
to superconductivity above $T_{\rm c}$ as earlier inferred for
CeIrIn$_5$ \cite{nair08}.

\section{X-ray diffraction}
\begin{figure}[t]
\center \includegraphics[width=8.6cm]{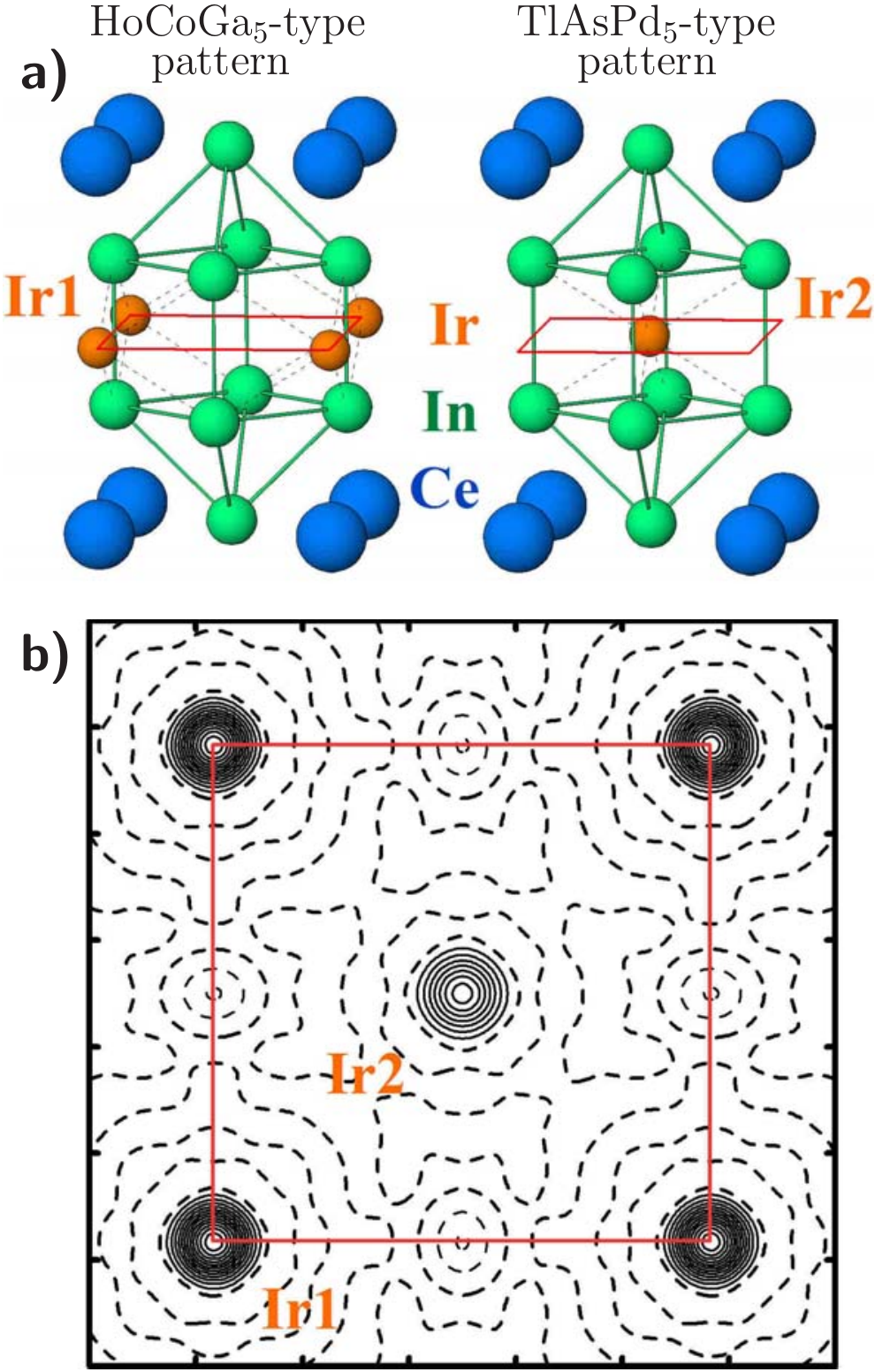} \caption{a) Two
closely related structure types HoCoGa$_5$ (left) and TlAsPd$_5$
(right). In the former structure, Ir occupies the position at $0\,
0\, \frac{1}{2}$ are occupied (Ir1). In comparison, the Ir2
position in the latter structure is shifted by $\langle
\frac{1}{2}\, \frac{1}{2}\, 0 \rangle$ within the $z =
\frac{1}{2}$ plane. b) Difference electron density calculated from
enhanced-resolution X-ray diffraction experiments on our single
crystals of CeIrIn$_5$. This indicates the presence of about 1\%
of Ir in Ir2 positions (TlAsPd$_5$ pattern) at $z = \frac{1}{2}$
and hence, a coexistence of the two structural patterns in the
crystal structure of CeIrIn$_5$ is inferred.} \label{fig1}
\end{figure}

Normal-resolution neutron diffraction reported earlier \cite{mosh}
(540 symmetry dependent reflections, $R$(F) = 0.051) indicated
that CeIrIn$_5$ crystallizes in the structure type HoCoGa$_5$
\cite{grin79}, {\it cf.} Fig.\ \ref{fig1}(a) left. However, the
results of the STM topography studies discussed below insinuated a
more complex structure, at least at the sample surface. In an
effort to possibly relate surface and bulk structural properties,
we performed enhanced-resolution diffraction experiments on
several samples CeIrIn$_5$ including those used for the STM
investigations discussed below. Our X-ray diffraction experiments
were conducted with a resolution comparable to the above-mentioned
neutron investigations (Mo\,K$\alpha$ radiation, 2$\theta_{\rm
max}$ = 53$^{\circ}$, 133 symmetry-independent reflections, $R$(F)
= 0.020). In a first approximation, these experiments confirmed
the crystal structure of the HoCoGa$_5$ type. Yet, recent
investigations of the coexistence of different structural
patterns in the modifications of TmAlB$_4$ \cite{mori07,yub09}
revealed the importance of the analysis of the difference electron
density maps even in the case of a low residual value $R$(F).
Analysis of 232 symmetry-independent, non-zero intensity ($I \geq
2\sigma (I)$) reflections up to $2\theta_{max} = 70^{\circ}$
(using Mo\,K$\alpha$ radiation) for CeIrIn$_5$ yielded $a$ =
4.6660(3)~\AA, $c$ = 7.5161(7) \AA\, and the subsequent refinement
resulted in $R$(F) = 0.024 within the space group $P$4/$mmm$. A
striking feature of this refinement was the fact that the
displacement parameter of Ir atoms (atomic number 77) is larger
than the one for the more `light' Ce (atomic number 58), which
suggests a lower occupation of the Ir position. Indeed, the
distribution of the difference electron density in the plane at $z
= 0.5$, calculated from these diffraction data and without Ir
atoms, exhibits maxima at the edges of the unit cell, see Fig.\
\ref{fig1}(b), positions marked Ir1. These maxima are expected for
the structural pattern of the HoCoGa$_5$ type, Fig.\ \ref{fig1}(a)
left. In addition, however, maxima of the difference electron
density were also found in the center of the unit cell, {\it i.e.}
at position Ir2, which is characteristic for the structure pattern
of TlAsPd$_5$ type \cite{elbo} (Fig.\ \ref{fig1}(a) right).
Further refinement resulted in occupancies of {\it occ}(Ir1) =
0.988(3) and {\it occ}(Ir2) = 0.012(3).

As symmetry-averaged data were used, the refinement based on the
unit weights required a fixed scale factor. The importance of the
scaling on the stability of the refinement was already pointed out
in the first structural study of CeIrIn$_5$ (Ref.\ 25). Because
the crystal structure of the investigated crystals of CeIrIn$_5$
reveals a non-negligible disorder of the Ir atoms, the
translational symmetry in strict sense is broken. Thus, we
performed a more elaborate refinement of the crystal structure
using all 1564 measured non-zero symmetry-dependent reflections
({\it i.e.} without averaging for symmetry equivalents). The
standard deviations for the refined parameters were calculated
using the number of the symmetry-independent reflections. In this
case, the refinement was stable without fixing the scale factor
(keeping goodness-of-fit, GOF, close to unity by an appropriate
scaling) and yielded the occupancies of {\it occ}(Ir1) = 0.989(3)
and {\it occ}(Ir2) = 0.011(3). Having in mind the importance of
the completeness of the diffraction data set ({\it i.e.} the
presence of {\it all} symmetrically equivalent reflections within
the given range of $\sin \theta / \lambda$), we performed the
final diffraction experiment on the same single crystal and
obtained the completeness  close to 100\%, applying a specifically
developed algorithm for collecting the diffraction data
\cite{wed14}. For the finally collected 1705 non-zero
symmetry-dependent and the 240 non-zero symmetry-independent
reflections (Mo\,K$\alpha$ radiation, $2\theta_{max} =
70^{\circ}$) the refinement yielded the occupancies of {\it
occ}(Ir1) = 0.988(2) and {\it occ}(Ir2) = 0.012(2) with $R$(F) =
0.023 and {\it occ}(Ir1) = 0.984(2) and {\it occ}(Ir2) = 0.016(2)
with $R$(F) = 0.016, respectively. As the occupation values
obtained in both cases agree within two estimated standard
deviations, these values were averaged for the final model (Ref.\
31\nocite{struc}). The presence of Ir atoms at two different
positions is a key observation for understanding the atomic
distribution on the surface as seen in the STM experiments.

Considering the similarities between CeIrIn$_5$ and CeCoIn$_5$ the
existence of structure pattern of TlAsPd$_5$ type may also be
expected in CeCoIn$_5$. The crystal structure of CeCoIn$_5$ was
originally studied using X-ray powder diffraction data
\cite{kal89} with no irregularities in the crystal structure
reported. However, our enhanced-resolution single crystal
diffraction experiment on CeCoIn$_5$ (Mo\,K$\alpha$ radiation,
$2\theta_{max} = 66.4^{\circ}$, 1001 observed symmetry-dependent
reflections) revealed occupancies of both possible Co positions
(Co1 at $0\, 0\, \frac{1}{2}$ and Co2 at $\frac{1}{2}\,
\frac{1}{2}\, \frac{1}{2}$) with the occupancy factors of {\it
occ}(Co1) = 0.985(8) and {\it occ}(Co2) = 0.015(8) (Ref.
33\nocite{struCo}). Despite the higher standard deviations caused
by a lower contribution of cobalt atoms to the overall diffraction
intensity, also here the refinement was stable including all
observed reflections when keeping the goodness-of-fit close to
unity by appropriate scaling.

\section{STM investigations of surface topography}
In an attempt to directly visualize the crystal structure as well
as the disorder discussed above we conducted topography
measurements by STM. Because STM is a particularly surface
\begin{figure}[t]
\includegraphics[width=8.4cm,clip=true]{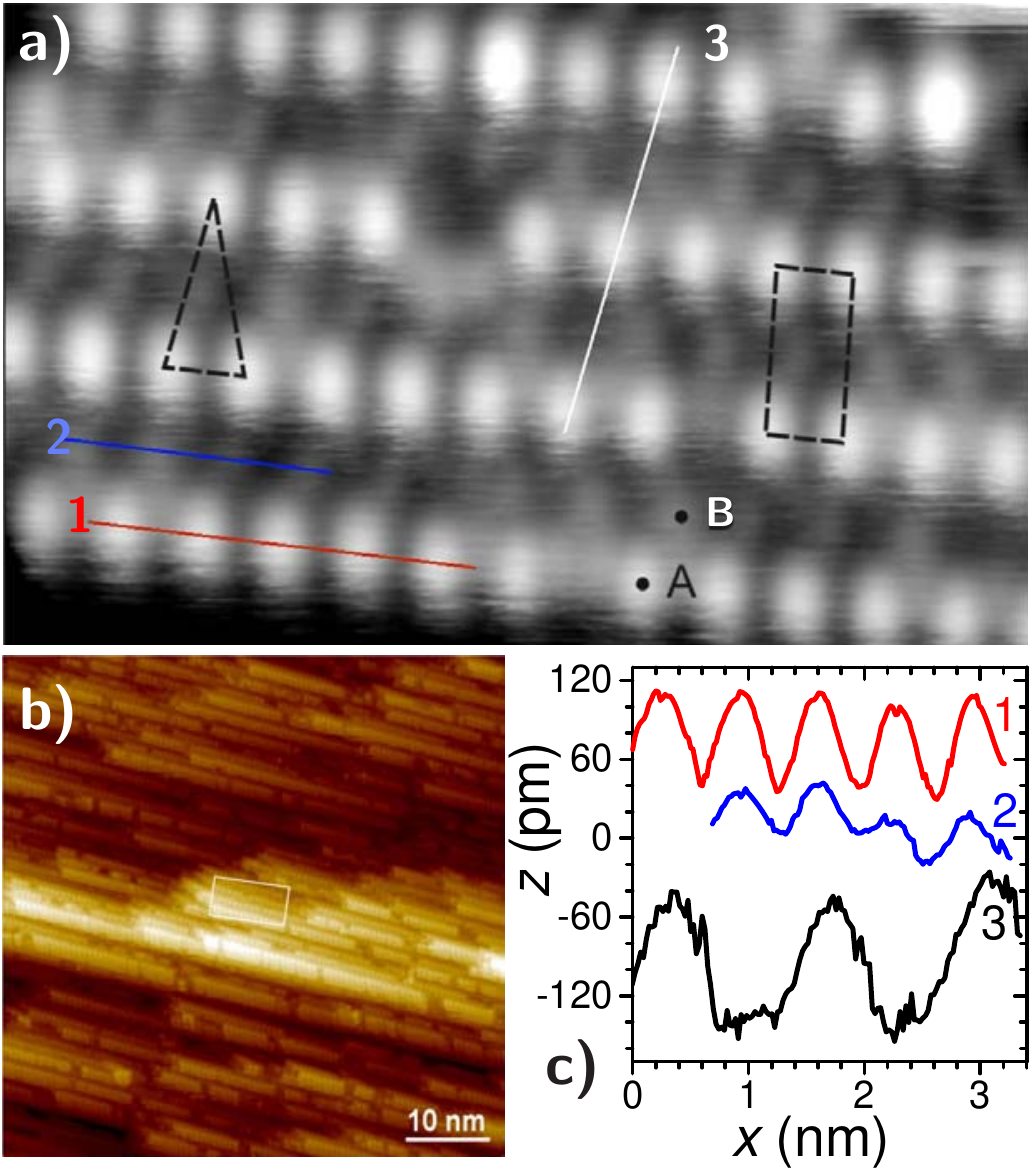}
\caption{a) STM topography on CeIrIn$_5$ after subtracting a plane
with an inclination of 37$^{\circ}$. The image covers an area of
$5.4\,\times\,8.6$ nm$^2$ and a total $z$-range of 0.63 nm. This
image was obtained at $V_g = +600$ mV, $I_{\rm set} =0.3$ nA, $T
=330$ mK. The markers {\sffamily\bfseries A} and
{\sffamily\bfseries B} indicate two kinds of atomic corrugations.
The dashed triangle and rectangle illustrate the two different
arrangements of atoms of type {\sffamily\bfseries A}. Lines 1 and
3 involve atoms of type {\sffamily\bfseries A}; line 2 those of
type {\sffamily\bfseries B}. b) Topography overview after
subtracting a plane of 37$^{\circ}$. The area shown in a) is
marked. c) Line scans indicated in a).} \label{atoms}
\end{figure}
sensitive technique special attention has to be paid with respect
to the sample surface quality. Therefore, the STM utilized here is
operated in UHV ($p \le 2 \times 10^{-9}$ Pa) and equipped for
{\it in situ} sample cleaving. Moreover, this STM can be operated
at sample temperatures as low as 0.3 K (allowing for an energy
resolution of $\le 100 \:\mu$eV) and in magnetic fields of up to
12 T. We stress again that in order to allow for a direct
comparison between the results obtained by XRD and STM identical
samples have been investigated.

Atomically resolved images of CeIrIn$_5$ were obtained within
areas of up to $60 \times 60$ nm$^2$, exhibiting various terraces
of up to a few ten nm in extent \cite{ernst09}, as exemplified in
Fig.~\ref{atoms}. In case of the overview topography presented in
Fig.~\ref{atoms}(b), a plane of inclination of 37$^{\circ}$ was
subtracted. An area within a terrace marked in Fig.~\ref{atoms}(b)
is magnified in Fig.~\ref{atoms}(a) and clearly shows atomic
resolution. The sample was mounted parallel to the
crystallographic $ab$-plane. Hence, the large value of the tilting
angle of 37$^{\circ}$ of the imaged sample area with respect to
the scanning plane points towards the fact that the terrace of
Fig.~\ref{atoms}(a) represents a lattice plane of low symmetry (we
emphasize that this plane correction was taken into consideration
when calculating the interatomic distances below). Note that this
tilting of the area presented in Fig.~\ref{atoms}(a) is maximum in
a direction perpendicular to scan lines 1 and 2. Line scans
\begin{figure}[t]
\includegraphics[width=8.4cm]{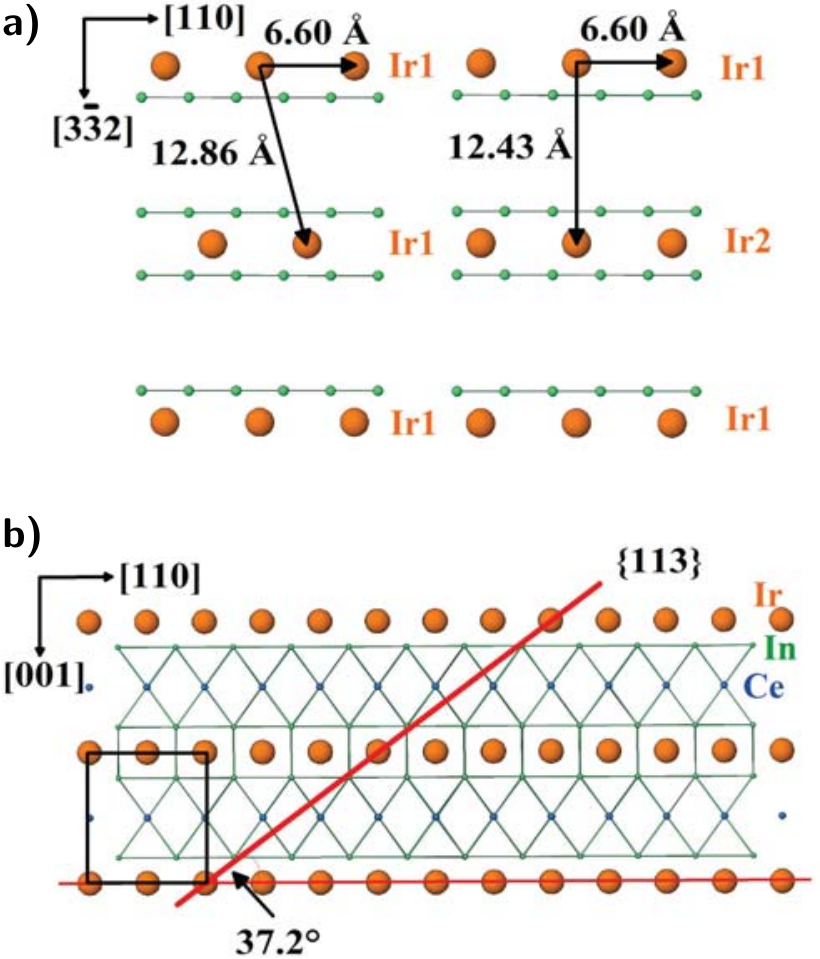}
\caption{a) Arrangement of the Ir and In atoms in the \{113\}
planes of the HoCoGa$_5$ (left) and the TlAsPd$_5$ structure
(right). b) Visualization of a \{113\} plane and its inclination
angle with respect to the (001) plane.}\label{arrange}
\end{figure}
through the prominent corrugations are presented in
Fig.~\ref{atoms}(c) and marked 1 and 3; the locations of these
scans are indicated in Fig.~\ref{atoms}(a). The distance between
the corrugations depends strongly on the direction, $\sim$6.7 \AA\
for line 1 and $\sim$13.7 \AA\ for line 3. Given these distances
it is unlikely that these prominent corrugations represent In
atoms (at least as long as there is no severe surface
reconstruction). The positive bias voltage $V_g = +600$ mV used
while acquiring the topography of Fig.~\ref{atoms}(a) may support
a more prominent visualization of Ir atoms since they accumulate
the strongest negative charge. Therefore, the most prominent
corrugations marked by {\sffamily\bfseries A} in
Fig.~\ref{atoms}(a) are probably Ir atoms. The corrugations marked
as {\sffamily\bfseries B} could then originate from the Ce atoms
of the intermittent Ce layers as shown in Figs.~\ref{arrange}(a)
and (b). This is corroborated by similar distances between the
corrugations of type {\sffamily\bfseries B} (line 2) and of type
{\sffamily\bfseries A} (line 1). Clearly, for corrugations of type
{\sffamily\bfseries B} the assignment to a certain atomic species
is even more problematic than for those of type
{\sffamily\bfseries A} because, in addition to the actual height,
also a changed density of states (DOS) may affect the apparent
height. However, in what follows a reversed assignment of
corrugations {\sffamily\bfseries A} to Ce and corrugations
{\sffamily\bfseries B} to Ir would not significantly change our
conclusions.

A tilting angle away from the \{001\} plane is of importance for
the possible identification of positions Ir1 and Ir2. Cleaves on
the related compound CeCoIn$_5$ resulted in \{001\} terminating
planes \cite{yazdani,davis13,yaz13}. However, within the \{001\}
plane the two positions Ir1 and Ir2 are difficult to distinguish
\begin{figure*}[t]
\centering \includegraphics[width=15.6cm,clip]{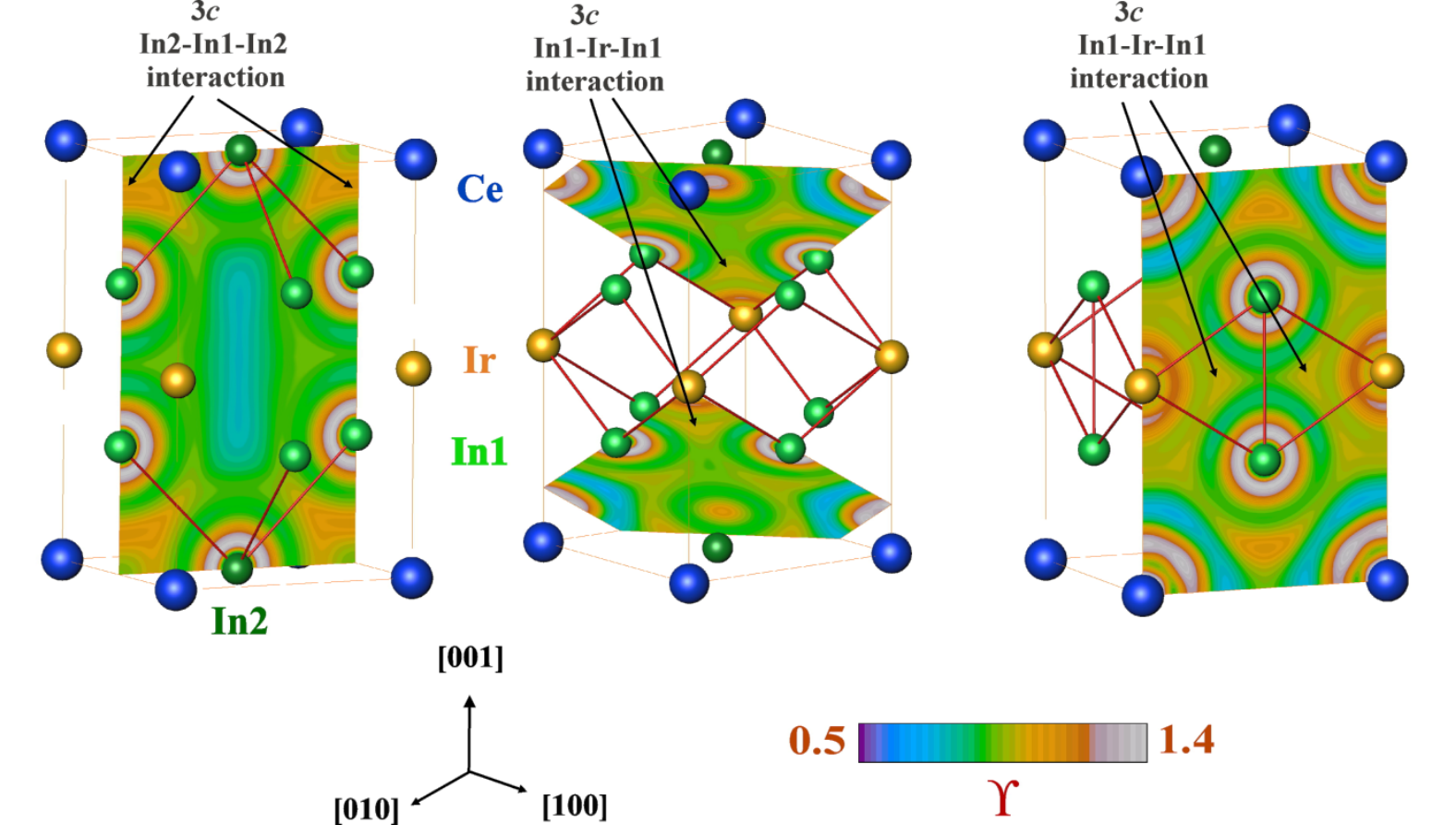}
\caption{Distribution of the Electron-Localizability-Indicator
(ELI, $\Upsilon$) in CeIrIn$_5$ revealing the 3$c$ interactions
within the anionic Ir-In framework, and mostly ionic interactions
of the Ce cations with the framework.} \label{ELI}
\end{figure*}
because of the aforementioned shift of Ir planes.

\section{Structural considerations}
The effective atomic charges in CeIrIn$_5$ were calculated
applying the QTAIM (Quantum Theory of Atoms in Molecules
\cite{bad94}) approach. As expected from the electronegativities
of the elements, Ir carries the largest negative charge ($-1.7$),
In atoms have charges close to zero ($+0.25$ for In1 and $-0.1$
for In2), and Ce sustains the largest positive charge of $+0.8$.
Taking this into account, the formula of the compound
should---from a chemical point of view---be rather written as
CeIn$_5$Ir; yet for historical reasons we continue to use the
`traditional' formula CeIrIn$_5$ (as well as CeCoIn$_5$) in this
work.

Further analysis of the chemical bonding in CeIrIn$_5$ (HoCoGa$_5$
type) applying the electron localizability approach \cite{wag}
reveals that the Ir and In atoms form three-dimensional framework
polyanions through the covalent (polar) three-center (3$c$)
interactions. Three independent kinds of such interactions are
observed. The first one (In2--In1--In2) is formed by In atoms only
within the CeIn$_3$ segment (Fig.\ \ref{ELI}, left), two others
(In1--Ir--In1) are found within the IrIn$_2$ segment (Fig.\
\ref{ELI}, middle and right). The last (sixth) shell of Ce is
absent suggesting an electron transfer to the Ir--In framework
(ionic interaction); the non-spherical
Electron-Localizability-Indicator (ELI, $\Upsilon$) distribution
in the fifth shell indicates that these electrons also participate
in the interaction within the valence region. Recent
investigations of the atomic arrangements at the surface of single
crystals of intermetallic compounds Al$_{13}$Co$_4$ (Ref.\ 38,
39\nocite{add09,shin11}) and Al$_{13}$Fe$_4$ (Ref.\
40\nocite{led13}) reveal that the terminating surface often
exhibits a large amount of covalent bonds.

Applying this approach to CeIrIn$_5$ we found that there are
likely three different crystallographic planes along which the
compound may preferentially cleave: the \{100\} planes containing
In2--In1--In2 bonds, and the \{001\} and \{113\} planes containing
In1--Ir--In1 interactions. Most likely for reasons of the bond
strength, in CeIrIn$_5$ the \{113\} planes are expected to be the
terminating ones upon cleaving, whereas in CeCoIn$_5$---showing
very similar atomic interactions---the \{001\} planes were found
\cite{davis13,yaz13,yazdani} as the terminating ones after
cleaving. Because the \{113\} planes are present in four different
orientations in the crystal, the cleavage of CeIrIn$_5$ may not be
regarded as similar to that in CeCoIn$_5$.

Consequently, we suggest that the terminating surface seen in
Fig.~\ref{atoms}(a) is a \{113\} plane. For such a plane an
inclination of 37.2$^{\circ}$ with respect to the \{001\} plane is
expected, {\it cf.} Fig.~\ref{arrange}(b), in good agreement with
the inclination of the surface observed in STM. Within this plane
the adjacent Ir atoms should be spaced by 6.6 \AA,
Fig.~\ref{arrange}(a). The distances of the {\sffamily\bfseries
A}-type corrugations within the lines ({\it e.g.} line scan 1 in
Fig.~\ref{atoms}(a)) observed in STM topography, ($6.7 \pm 0.3$)
\AA, are in reasonable agreement with the shortest distances ({\it
i.e.} along the $[$110$]$ direction) between the Ir atoms within
the \{113\} plane of CeIrIn$_5$. Also, the nearest Ce atoms within
this plane have the same distances which is consistent with the
observation of analogous distances of corrugations
{\sffamily\bfseries A} and {\sffamily\bfseries B} in
Fig.~\ref{atoms}(a). Within the \{113\} plane and perpendicular to
the $[$110$]$ direction, the next row of Ir (or Ce) atoms is
located 12.43 \AA\ away, again in good agreement with our
observation, {\it cf.} line scan 3 in Figs.~\ref{atoms}(a) and
(c). Here, however, there is an obvious difference between the
HoCoGa$_5$ and TlAsPd$_5$ structure types: Within the HoCoGa$_5$
structure pattern the Ir atoms are shifted along $[$110$]$ such
that the Ir atoms form isosceles triangles, as shown in
Fig.~\ref{arrange}(a), left. In contrast, the TlAsPd$_5$ structure
pattern result in a rectangular arrangement of the Ir atoms, see
Fig.~\ref{arrange}(a), right. Clearly, both structure types are
observed within the area of Fig.~\ref{atoms}(a) as indicated by
the dashed triangle and rectangle. Scan line 3 correspondingly
follows a HoCoGa$_5$ structural pattern; the observed distance
between corrugations of (13.7 $\pm$ 0.8) \AA\ agrees nicely with
the expected Ir distance of 12.86 {\AA}, see
Fig.~\ref{arrange}(a), left. Scan line 3 not only shows the
prominent corrugations {\sffamily\bfseries A} but also the atoms
of type {\sffamily\bfseries B} can be suspected.

Although the triangular arrangement of atoms {\sffamily\bfseries
A} is much more dominant, we lack sufficient statistics in our STM
topography to estimate the frequency of the two structure types
and compare with to the result from XRD experiments. Moreover, the
cleaving might have taken place along planes of increased defect
density which may render a quantitative comparison between
bulk-sensitive X-ray and surface-sensitve STM measurements
difficult. The extent of the rectangular arrangement corresponding
to the TlAsPd$_5$ structure pattern is, according to our STM
topography, likely in the order of a few atomic distances. One
might then speculate that the pattern of HoCoGa$_5$ structure type
observed here are related to the structural imperfections which
may cause the discrepancy between bulk and resistive
superconducting transition temperature in CeIrIn$_5$ \cite{bia01}.
We note here that, although the total volume fraction of the
impurity phase, {\it i.e.} the TlAsPd$_5$ structure pattern, is
rather small, the physical properties might be influenced within a
larger volume.

Recently, the intergrowth of $\alpha$- and $\beta$-type YbAlB$_4$
single crystals was studied \cite{yub13}. The latter compound is a
heavy fermion superconductor exhibiting quantum criticality
\cite{nak08}, with a specific heat coefficient being more than
twice as large as in the $\alpha$-type compound \cite{mac07}. In
$\beta$-type TmAlB$_4$ a significantly enhanced magnetic
interaction was found \cite{mor10} if compared to its
$\alpha$-type counterpart. These observations again underline the
importance of detailed structural investigations for shedding
light on intricate physical properties. We attempted to conduct
Scanning Tunneling Spectroscopy (STS) within the atomically
resolved surface areas of our CeIrIn$_5$ to possibly identify
differences in the physical properties of the HoCoGa$_5$ and
TlAsPd$_5$ structure types. However, no significant differences
were found. Specifically, no indication for the opening of a
superconducting gap was observed in the measured tunneling
conductance within neither areas, independent of whether they
belong to the HoCoGa$_5$ or the TlAsPd$_5$ structure types.
Therefore, possible differences in the superconducting properties
of the two different structure types could, unfortunately, not
directly be visualized. We speculate that superconductivity may be
diminished at the sample surface. We also note that our
measurement base temperature $T =$ 0.32 K is just below the bulk
$T_{\rm c} \approx$ 0.4 K.

\section{Precursor state to superconductivity}
For CeCoIn$_5$ with $T_{\rm c} \approx$ 2.3 K the situation is
much more comfortable, and the formation of heavy fermions and of
superconductivity was investigated by STS
\cite{yazdani,davis13,yaz13}. As mentioned above, however, all our
attempts to cleave (more than 30) and those reported
\cite{yazdani,davis13,yaz13} resulted in surfaces with terraces of
\{001\} orientation rendering the observation of a possible
similar TlAsPd$_5$ structure pattern in CeCoIn$_5$ highly
difficult. Consequently, a direct visualization of the two
different structure types as found for CeCoIn$_5$ by XRD
investigations was not possible.

Earlier magnetotransport investigations on the system CeIrIn$_5$
indicated the existence of a precursor state to superconductivity
\cite{nair08}. Moreover, it was demonstrated that the Hall
coefficient $R_{\rm H}$ and the magnetoresistance $\varrho_{\rm
xx}$ are governed by two distinct scattering times \cite{nair09}.
Both observations are reminiscent of the behavior found for the
copper oxide superconductors and are consistent with a scenario in
which incipient antiferromagnetic fluctuations crucially influence
\begin{figure}[t]
\includegraphics[width=8.6cm,clip]{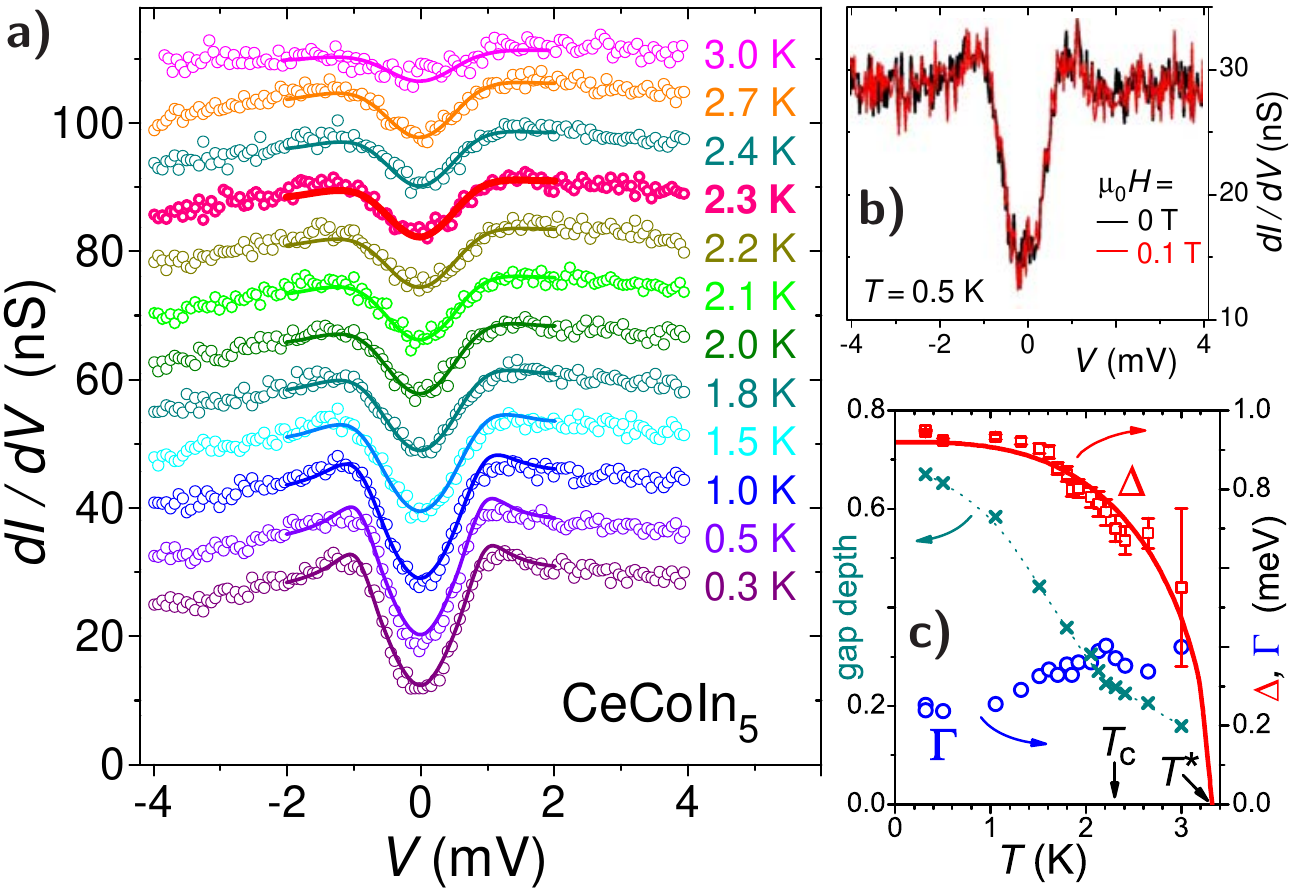}
\caption{Tunneling spectroscopy on CeCoIn$_5$. a) Measured
tunneling conductance $g(V) = dI / dV$ (open circles, set point
parameters: $V$ = +14 mV, $I_{\rm set}$ = 340 pA), for clarity
shifted vertically (except the one obtained at $T =$ 0.3 K). The
lines are results of the fit at each temperature (see text). b)
Comparison of $g(V) = dI / dV$ for zero field and $\mu_0 H =$ 0.1
T. c) Values of $\Delta (T)$ and $\Gamma (T)$ resulting from the
fits of the tunneling conductance at each measured temperature in
a). Also shown is the relative depth of the gap (see text).}
\label{fig-sts}
\end{figure}
the magnetotransport in both classes of materials, the fermion
systems as well as the cuprates. In case of the cuprate
superconductors, STS has proven to be a powerful tool for the
investigation of the superconducting gap and, specifically, of the
pseudogap \cite{fis07}. Therefore, we investigated CeCoIn$_5$ by
STS \cite{ernst09} with focus on a possible precursor state
similar to the one found in CeIrIn$_5$.

In Fig.~\ref{fig-sts}(a), differential tunneling conductance
($g(V) = dI/dV$) spectra are presented as obtained for CeCoIn$_5$
within atomically flat terraces and within a temperature range
0.32 K $\le T \le$ 3 K. Upon increasing temperature the zero-bias
conductance $g(V=0)$ increases, indicating the expected closing of
the gap. The gap, however, does not disappear at $T_{\rm c}
\approx$ 2.3 K (bold red markers), but is still clearly visible at
$T$ = 3 K. We note that the observed gap is genuine in that the
application of a small magnetic field of $\mu_0 H =$ 0.1 T
perpendicular to the sample surface does not alter the gap
spectrum notably, see Fig.~\ref{fig-sts}(b). The observation of
coherence peaks at lowest temperatures may indicate that these gap
structures are indeed related to superconductivity, rather than to
Kondo interactions \cite{yazdani,ernst11}. This assessment is
corroborated by the temperature dependence of the relative depth
of the gap, Fig.\ \ref{fig-sts}(c), which does not exhibit the
logarithmic decay expected \cite{cos00} for Kondo systems. Here,
the gap depth was taken as the difference in conductance $g(V)$ at
the coherence peaks ({\it i.e.} around $|V| \approx 1.1$ mV) and
at $V =$ 0, normalized to $g(V\! =\! -4\,\rm{mV})$ ({\it i.e.}
away from the gap structure) after subtracting a linear
background.

Assuming a $d_{x^2-y^2}$ symmetry of the superconducting order
parameter as suggested elsewhere \cite{iza01,vor06,davis13,yaz13},
the tunneling density of state (which is directly related to $g(V)
= dI / dV$) within the BCS framework is given by \cite{won94}
\begin{equation}
\rho(E)\propto\mbox{Re}\int_0^{2\pi}\frac{d\phi}{2\pi}
\frac{E-i\Gamma}{\sqrt{(E-i\Gamma)^2-\Delta^2\cos^2(2\phi)}}\; .
\label{dos}
\end{equation}
Here, $\Delta$ is the maximum value of the angular dependent gap
function and $\Gamma$ is an additional lifetime broadening
parameter. For each measured temperature the resulting fit is
included as line in Fig.~\ref{fig-sts}(a) while the values of
$\Delta$ and $\Gamma$ are presented in Fig.~\ref{fig-sts}(c). As
expected, $\Gamma$ shows only little temperature dependence (its
low-temperature value of 0.25 meV is in accord with results from
point contact spectroscopy \cite{greene05}) while $\Delta$
decreases with temperature. For nodal superconductors, the
temperature dependence of the order parameter can be approximated
\cite{dora01} by $\Delta(T)=\Delta_0\sqrt{1-(T/T^*)^3}$. The red
line in Fig.~\ref{fig-sts}(c) represents a corresponding fit of
our results which yields $T^* \approx$ 3.3~K as the temperature at
which the order parameter vanishes.

The zero-temperature value of the order parameter $\Delta_0
\approx$ 0.93 meV results in $2 \Delta_0 / k_{\rm B} T_{\rm c}
\approx$ 9.3. Although such large values have been reported
\cite{goll03,rou05}, smaller numbers \cite{greene08} of $\sim$6
appear more reliable. A range of $\Delta_0$ values was suggested
to result from the quality of the sample surface \cite{sum08}.
Indeed, recent STM investigations \cite{yaz13} showed the
development of a pseudogap-like feature just above 5 K, but only
on one type of surface (for the superconducting gap at lower
temperature, $\Delta_{\rm SC} \approx$ 0.6 meV was reported
\cite{davis13,yaz13}). Note that our data do not provide any
indication for multiple order parameters \cite{rou05}.

In the following we focus on the appearance of a pseudogap-like
feature above $T_{\rm c}$ which is estimated to dissolve around
3.3 K. A pseudogap refers to a metallic regime within the
temperature range $T_{\rm c} < T < T^{*}$ where the DOS is
considerably suppressed at least in parts of the Brillouin zone
\cite{tim99}. At least for the cuprate superconductors, the
$d$-wave nature of the order parameter holds for the
superconducting and the pseudogap regime \cite{tim99,val06}
suggesting that the above defined $T^{*}$ (vanishing order
parameter) equivalently limits the pseudogap regime towards higher
temperatures. Pseudogaps have been observed in a number of systems
displaying electronically mediated superconductivity. The
pseudogap regime was first discovered in hole-doped cuprates where
NMR measurements pointed to pseudo-gapped spin excitations
\cite{war89}. It subsequently was seen by photoemission and
detected in transport measurements (see {\it e.g.} Ref.\ 58 for
details on the pseudogap regime in the cuprates). Similar
pseudogap regimes have also been seen in other unconventional
superconductors belonging to the iron pnictide family \cite{xu11}
and the organic transfer salt superconductors \cite{kang11}.

As the pseudogap feature evolves smoothly out of the
superconducting regime it is tempting to suspect an analogy to the
precursor state to superconductivity found \cite{nair08} in
CeIrIn$_5$ as well as the pseudogap reported \cite{kaw05} for
CeRhIn$_5$ in a pressure regime where antiferromagnetism and
superconductivity coexist. A pseudogap-like feature was speculated
to exist in CeCoIn$_5$ up to 3.3 K at ambient pressure based on
measurements of electrical resistivity under pressure
\cite{Sidorov}. Also, the fourfold anisotropy expected for a
$d$-wave superconductor was observed up to 3.2 K \cite{iza01}.

An interpretation of the pseudogap formation as a precursor
phenomenon to unconventional superconductivity may appear natural
as the suppression of the DOS at temperatures $T_{\rm c} < T <
T^{*}$ reduces the kinetic energy increase that accompanies the
opening of a full charge excitation gap at $T_{\rm c}$. Indeed, a
suppression of the normal-state DOS is a consequence of incoherent
pairing above $T_{\rm c}$, a phenomenon which is more pronounced
in superconductors with short coherence lengths. $T^{*}$ then
marks the crossover energy scale to a regular metallic DOS. Yet,
the origin and nature of the $T^{*}$-line, and concomitantly the
pseudogap regime in (primarily) the cuprates has remained
controversial \cite{mil06}. According to some theories, $T^{*}$ is
associated with a sharp phase transition into an ordered state out
of which superconductivity arises. Competing interactions, {\it
e.g.} superconductivity and charge density wave order seem quite
generally to promote the suppression of the DOS above $T_{\rm c}$.
A recent numerical study of the single-band Hubbard model based on
the cluster dynamical mean field theory with clusters containing
up to 16 sites finds that the pseudogap regime itself competes
with superconductivity and that the superconductivity in this
model tends to reduce the charge excitation gap when it emerges
from the pseudogap phase \cite{gull13}. The pseudogap phase in the
Hubbard model separates a Mott insulator from the superconducting
ground state.

The pseudogap regime in CeCoIn$_5$ thus seems to be of a different
origin than the one in the cuprates as superconductivity in
CeCoIn$_5$ does not arise near a Mott transition, but instead in
close proximity of a spin-density wave quantum critical point. The
coherence length $\xi$ in CeCoIn$_5$ is only about 48 \AA
\cite{yaz13,tay02}, so that the pseudogap regime may rather be
caused by incoherent pairing. The situation is different for
CeRhIn$_5$, where superconductivity is found near the onset of
magnetism that is characterized by a critical Kondo destruction
\cite{park06}. This local quantum criticality is associated with a
Mott-like localization of the 4$f$-moments \cite{si01}. The
observation \cite{kaw05} of a pseudogap regime accompanying
superconductivity in CeRhIn$_5$ thus helps to elucidate the role
kinetic energy gain at $T_{\rm c}$ plays in unconventional
superconductors. For spin-fluctuation mediated superconductors
this question has recently been raised in the context of
CeCu$_2$Si$_2$ \cite{sto11,sto12}. A better understanding of the
occurrence and origin of the pseudogap regime in heavy fermion
superconductors may thus foster a deeper understanding of the
intricate interplay between superconductivity and magnetism that
apparently underlies almost all electronically driven
superconductors.

\section{Conclusion}
The investigated single crystals of CeIrIn$_5$ exhibited a
coexistence of the expected HoCoGa$_5$-type majority structural
pattern with a minority pattern (1.2\%) of the TlAsPd$_5$ type.
Here, the combination of different techniques proved essential:
While the XRD measurements showed that these structural impurities
are present throughout the bulk of the samples (not just at their
surfaces) the STM investigations directly visualized these latter,
additional structural patterns at the surface. Clearly, such
``structural impurities'' are to be expected whenever very closely
related structure types can be realized within a given compound,
and should be borne in mind in the consideration of complex
physical properties.

\section*{Acknowledgements}
This work is partially supported by the German Research Foundation
through DFG Forschergruppe 960. Z.F. acknowledges support through
NSF-DMR-0801253. Work at Los Alamos was performed under the
auspices of the US DOE, Office of Basic Energy Sciences, Division
of Materials Sciences and Engineering.


\begin{thebibliography}{99}
\bibitem{struc} Structural data for CeIrIn$_5$: space group
P4/$mmm$; $a$ = 4.6660(3) \AA, $c$ = 7.5161(7) \AA; 1Ce in 1($a$)
$0\, 0\, 0$, $B_{eq}$ = 0.65(1); 4In1 in 4($i$) 0 $\frac{1}{2}$
0.30535(7), $B_{eq}$ = 0.95(1); 1In2 in 1($c$) $\frac{1}{2}
\frac{1}{2} 0$, $B_{eq}$ = 0.9(2); 0.986(2)Ir1 in 1($b$) 0 0
$\frac{1}{2}$, $B_{eq}$ = 0.74(1); 0.014(2)Ir2 in 1($d$)
$\frac{1}{2}\, \frac{1}{2}\, \frac{1}{2}$, $B_{iso}$ = 0.8(1);
$R$(F) = 0.016 for 240 observed symmetrically independent
reflections; $B_{eq}$ = 1/3[$B_{11} a^{*2} a^2 + \ldots 2B_{23}
b^* c^* b c \cos \alpha$]; $B_{iso}$ stands for the isotropic
displacement parameter of Ir2, all other atoms were refined in
anisotropic approximation of the atomic displacement.
\bibitem{struCo} Structural data for CeCoIn$_5$: space group
P4/$mmm$; $a$ = 4.6132(5) \AA, $c$ = 7.748(1) \AA; 1Ce in 1($a$)
$0\, 0\, 0$, $B_{eq}$ = 0.73(2); 4In1 in 4($i$) 0 $\frac{1}{2}$
0.3095(1), $B_{eq}$ = 1.04(2); 1In2 in 1($c$) $\frac{1}{2}
\frac{1}{2} 0$, $B_{eq}$ = 0.94(3); 0.985(8)Co1 in 1($b$) 0 0
$\frac{1}{2}$, $B_{eq}$ = 0.74(1); 0.015(8)Co2 in 1($d$)
$\frac{1}{2}\, \frac{1}{2}\, \frac{1}{2}$, $B_{iso}$ = 0.8(1);
$R$(F) = 0.032 for 1001 observed reflections, refinement including
the symmetrically independent reflections only was instable;
$B_{eq}$ = 1/3[$B_{11} a^{*2} a^2 + \ldots 2B_{23} b^* c^* b c
\cos \alpha$]; $B_{iso}$ stands in this case for the isotropic
displacement parameter of Co2, all other atoms were refined in
anisotropic approximation of the atomic displacement.
\end{thebibliography}
\end{document}